\documentclass[preprint]{aastex}
\usepackage{amssymb,amsmath,graphicx,epstopdf,natbib,upgreek}

\begin{document}

\title{Simulating Radiative Magnetohydrodynamical Flows with AstroBEAR: Implementation and Applications of Non-equilibrium Cooling}
\author{E. C. Hansen \altaffilmark{1}, P. Hartigan \altaffilmark{2},  A. Frank \altaffilmark{3}, A. Wright \altaffilmark{2}, and J. C. Raymond \altaffilmark{4}}
\altaffiltext{1}{Laboratory for Laser Energetics, University of Rochester, Rochester, NY 14623, USA}
\altaffiltext{2}{Department of Physics and Astronomy, Rice University, 6100 S. Main, Houston, TX 77521, USA}
\altaffiltext{3}{Department of Physics and Astronomy, University of Rochester, Rochester, NY 14627, USA}
\altaffiltext{4}{Harvard-Smithsonian Center for Astrophysics, 60 Garden Street, Cambridge, MA 02138, USA}
\email{ehan@lle.rochester.edu}


\newcommand{\HI}{\mbox{H\,{\footnotesize I}}}         
\newcommand{\HII}{\mbox{H\,{\footnotesize II}~}}       
\newcommand{\Hions}{\mbox{H\,{\footnotesize I-II}}}         
\newcommand{\HeI}{\mbox{He\,{\footnotesize I}}}       
\newcommand{\HeII}{\mbox{He\,{\footnotesize II}}}     
\newcommand{\HeIII}{\mbox{He\,{\footnotesize III}}}   
\newcommand{\Heions}{\mbox{He\,{\footnotesize I-III}}}       
\newcommand{\Heionsii}{\mbox{He\,{\footnotesize I-II}}}       
\newcommand{\SI}{\mbox{S\,{\footnotesize I}}}         
\newcommand{\SII}{\mbox{S\,{\footnotesize II}}}       
\newcommand{\SIII}{\mbox{S\,{\footnotesize III}}}     
\newcommand{\SIV}{\mbox{S\,{\footnotesize IV}}}       
\newcommand{\Sions}{\mbox{S\,{\footnotesize II-IV}}}       
\newcommand{\Sionsv}{\mbox{S\,{\footnotesize II-V}}}       
\newcommand{\Sionsvv}{\mbox{S\,{\footnotesize I-V}}}       
\newcommand{\Sionsxvii}{\mbox{S\,{\footnotesize V-XVII}}}       
\newcommand{\NI}{\mbox{N\,{\footnotesize I}}}         
\newcommand{\NII}{\mbox{N\,{\footnotesize II}}}       
\newcommand{\NIII}{\mbox{N\,{\footnotesize III}}}       
\newcommand{\Nions}{\mbox{N\,{\footnotesize I-II}}}         
\newcommand{\Nionsv}{\mbox{N\,{\footnotesize I-V}}}         
\newcommand{\Nionsviii}{\mbox{N\,{\footnotesize III-VIII}}}         
\newcommand{\OI}{\mbox{O\,{\footnotesize I}}}         
\newcommand{\OII}{\mbox{O\,{\footnotesize II}}}       
\newcommand{\OIII}{\mbox{O\,{\footnotesize III}}}       
\newcommand{\Oions}{\mbox{O\,{\footnotesize I-II}}}         
\newcommand{\Oionsv}{\mbox{O\,{\footnotesize I-V}}}         
\newcommand{\Oionsviii}{\mbox{O\,{\footnotesize III-VIII}}}         
\newcommand{\SiII}{\mbox{Si\,{\footnotesize II}}}       
\newcommand{\Siionsxiv}{\mbox{Si\,{\footnotesize II-XIV}}}       
\newcommand{\FeII}{\mbox{Fe\,{\footnotesize II}}}       
\newcommand{\Feions}{\mbox{Fe\,{\footnotesize II-XXVI}}}       
\newcommand{\MgII}{\mbox{Mg\,{\footnotesize II}}}       
\newcommand{\Mgions}{\mbox{Mg\,{\footnotesize I-XII}}}       
\newcommand{\CII}{\mbox{C\,{\footnotesize II}}}       
\newcommand{\Cions}{\mbox{C\,{\footnotesize II-IV}}}       
\newcommand{\Cionsv}{\mbox{C\,{\footnotesize I-V}}}       
\newcommand{\Cionsvi}{\mbox{C\,{\footnotesize I-VI}}}       
\newcommand{\Neionsv}{\mbox{Ne\,{\footnotesize I-V}}}       
\newcommand{\Neionsxi}{\mbox{Ne\,{\footnotesize II-XI}}}       

\begin{abstract}
Radiative cooling plays a crucial role in the dynamics of many astrophysical flows, and is particularly important in the dense shocked gas within Herbig-Haro (HH) objects and stellar jets.
Simulating cooling processes accurately is necessary to compare numerical simulations with existing and planned observations of HH objects, such as those from the Hubble Space Telescope and the James Webb Space Telescope.
In this paper we discuss a new, non-equilibrium cooling scheme we have implemented into the 3-D magnetohydrodynamic (MHD) code AstroBEAR.
The new cooling function includes ionization, recombination, and excitation of all the important atomic species that cool below 10000 K. 
We tested the routine by comparing its predictions with those from the well-tested 1-D Cox-Raymond shock code \citep{Raymond79}.
The results show that AstroBEAR accurately tracks the ionization fraction, temperature, and other MHD variables for all low-velocity ($\lesssim$ 90 km/s) magnetized radiative shock waves.
The new routine allows us to predict synthetic emission maps in all the bright forbidden and permitted lines observed in stellar jets, including H$\alpha$, [\NII], [\OI], and [\SII].
We present an example as to how these synthetic maps facilitate a direct comparison with narrowband images of HH objects.
\end{abstract}


\section{Introduction}
\label{sec:ch1intro}
Radiative processes such as absorption, emission, and scattering are prevalent in astrophysical gases.
Observations of such gases are made possible by the radiation that escapes and reaches the observer.
This emission is often an integral part of a system as it is an energy loss (i.e. cooling) and can affect the evolution of the gas.
Star forming regions, supernova remnants, planetary nebulae, and \HII regions are a few examples of systems where such cooling is important because the cooling timescales can be much shorter than the hydrodynamical timescales.

Much of the gas in these systems is H and He, which cool inefficiently at low temperatures ($\leq$ 10\textsuperscript{3} K) because their lowest excitation levels occur at relatively high energies.
In other words, the Boltzmann factor ($\exp{(\frac{-E}{kT})}$) is small at low temperatures when the gas is neutral and in ionization equilibrium.
However, when the gas is shocked, the sudden increase in temperature can lead to cooling rates that are over two orders of magnitude higher than the equilibrium value.
Atomic processes such as ionization and recombination also become important when the gas is not in ionization equilibrium.
The combination of shock heating and strong cooling can lead to a wide range of temperatures and ionization fractions, which is why a careful treatment of non-equilibrium cooling is important when shock waves are present.
One regime in which such processes play a key role is in the outflows from young stellar objects (YSOs).
These outflows are observed as Herbig-Haro (HH) jets, and they contain many shocks and heterogeneous cooling regions.
Both the shocks and cooling regions are readily observed in optical emission lines with instruments such as the Hubble Space Telescope (HST).

The motivation of this work comes from time-resolved HST observations of HH objects \citep[see e.g.][]{Hartigan11}.
In numerical simulations, non-equilibrium cooling is necessary to accurately predict the emission from an HH object.
The ideal magnetohydrodynamic (MHD) equations with cooling coupled with ionization and recombination equations for various species can result in a better estimate of the temperature and ionization state of the gas.
With such models, it is possible to produce synthetic emission maps, and in HH objects, emission is observed in H$\alpha$ (shock diagnostic) and [\SII] (post-shock cold flow diagnostic) \citep{Heathcote96}.


As we describe in this paper, AstroBEAR calculates the non-equilibrium rate equations for each bright emission line in the cooling zone with a separate code, and these cooling rates combine to form a look-up table for the total cooling rate.
That rate depends solely upon the ionization fraction of H, the electron density, and the electron temperature, parameters that AstroBEAR follows in each time step within the simulation. 
This scheme works well below 10\textsuperscript{4} K because the ionization states of all the strong coolants are either fixed or are tied to the ionization state of H through strong charge exchange reactions.
At higher temperatures, the code takes forbidden line cooling from the Dalgarno-McCray cooling curves \citep{Dalgarno72}, but calculates non-equilibrium recombination cooling from H and He explicitly in each step.
Our code's ability to accurately track the temperature and ionization state of the gas makes it possible to produce synthetic emission maps of various emission lines.
Previous studies of outflows using AstroBEAR's new cooling method have already been done \citep{Hansen15a,Hansen17}, and synthetic emission maps of H$\alpha$ and [\SII] were presented in those works.

Many other hydrodynamic codes calculate non-equilibrium cooling, and we list several of these codes here: YGUAZU-A \citep{Raga00}, FLASH \citep{Fryxell00}, PLUTO \citep{Tesileanu08}, ENZO \citep{Bryan14}, NIRVANA \citep{Ziegler16}, and RAMSES \citep{Teyssier02}.
All of these codes have been implemented in 3-D and have support for magnetic fields.
The implementation of non-equilibrium cooling in each of these codes is different, and each code may contain additional multi-physics such as radiation transfer, self-gravity, heat conduction, magnetic resistivity, etc.
Thus, the choice of which code to use depends on the problem of interest and personal preference.
Table~\ref{tab:codes} lists these codes and which ionic species they use to calculate cooling rates.

The species listed in Table~\ref{tab:codes} are separated into two columns (``Non-equilibrium'' and ``Equilibrium'').
In the ``Equilibrium'' column, these species contribute to the overall cooling but with ionization equilibrium assumed.
In the ``Non-equilibrium'' column, species are either explicitly tracked or charge-exchange is used to obtain contributions to cooling without assuming ionization equilibrium (i.e., non-equilibrium).
It is important to explicitly track H and He to correctly predict hydrodynamic quantities such as density and temperature, however, it is not always necessary to explicitly track the heavier elements (i.e., metals).
To predict emission line fluxes accurately, species should be treated in the non-equilibrium regime.
In AstroBEAR, the framework exists for future users of the code to implement explicit tracking of species (e.g., \OIII) if other emission lines are desired.

\begin{deluxetable}{cc p{150pt}|p{130pt}}
    \tablenum{1}
    \tablecolumns{4}
    \tablewidth{0pt}
    \tablecaption{Capabilities and species included in cooling function for various hydrodynamic codes \label{tab:codes}}
    \tablehead{ & \colhead{Charge} & \multicolumn{2}{c}{\hspace{20pt}Species} \\
    \colhead{Code} & \colhead{Exchange} & \colhead{Non-equilibrium} & \colhead{Equilibrium}}
    \startdata
        AstroBEAR & Y & \Hions, \Heions, \Nions, \Oions, \Sions & \Cionsvi, \Nionsviii, \Oionsviii, \Neionsxi, \Sionsxvii, \Mgions, \Siionsxiv, \Feions \\
        YGUAZU-A & Y & \Hions, \Cions, \Nionsv, \Oionsv, \Neionsv, \Sionsv & \multicolumn{1}{c}{none} \\
        FLASH & Y & \multicolumn{1}{c}{varies} & \multicolumn{1}{c}{varies} \\
        PLUTO & Y & \Hions, \Heionsii, \Cionsv, \Nionsv, \Oionsv, \Neionsv, \Sionsvv & \MgII, \SiII, \FeII \\
        ENZO & N & \Hions, \Heions, D, D\textsuperscript{+}, H$_2$, H$_2$\textsuperscript{+}, HD, HD\textsuperscript{+} & \multicolumn{1}{c}{none} \\
        NIRVANA & N & \Hions, \Heions, H\textsuperscript{-}, H$_2$, H$_2$\textsuperscript{+} & \multicolumn{1}{c}{none} \\
        RAMSES & N & \Hions, \Heions & \multicolumn{1}{c}{none}
    \enddata
\end{deluxetable}

Many of the codes treat non-equilibrium radiative cooling in a similar fashion.
YGUAZU-A \citep{Raga00} uses the same cooling as its predecessor code CORAL \citep{Raga97}.
YGUAZU-A/CORAL uses a network of 18 different rate equations for various ionic species to obtain a non-equilibrium cooling function.
AstroBEAR also uses rate equations (for H, He, and S), but it uses table look-ups for the other species.
Both codes are appropriate for approximate temperature ranges between 10\textsuperscript{3} and 10\textsuperscript{5} K.
The codes track different species, so the preferred code also depends on which emission lines are desired for analysis.
CORAL requires an assumption of axisymmetry, but YGUAZU-A can be run in 3-D.

Magnetic fields may be important, especially in cooling regions behind shock fronts.
For typical ISM conditions, the thermal and magnetic pressures are comparable, but at shock fronts the thermal pressure usually dominates.
In cooling regions, the magnetic field is compressed along with the gas, thus the magnetic pressure can begin to dominate \citep{Hartigan15}.
MHD codes like AstroBEAR are better-suited for simulating regions where fields may play an important role.

FLASH is an open source code, and different users have incorporated different forms of cooling to suit their needs.
For example, \citet{Gaspari11} used a metallicity dependent cooling function in FLASH which was appropriate for cooling in the intercluster medium (ICM) above 10\textsuperscript{4} K.
With an assumed metallicity, the cooling rate was reduced to a function dependent only on temperature based on cooling functions from \citet{Sutherland93}.
This treatment is different from AstroBEAR, but, in this case, FLASH is used in a completely different regime: the ICM.
The current version of non-equilibrium cooling in AstroBEAR is more appropriate for higher density and lower temperature regions such as the interstellar medium (ISM).

PLUTO has been used to study radiative shocks in YSO jets \citep{Tesileanu14}.
Similar to the other aforementioned codes, PLUTO tracks several ionic species to calculate collisionally excited line radiation in the optically thin limit.
Like AstroBEAR, it adds cooling contributions from \MgII, \SiII, and \FeII ~without explicitly tracking their number densities.
PLUTO also uses look-up tables for more efficient computation.
One can see from Table~\ref{tab:codes} that the PLUTO code has a relatively large network of ionic species, but we will show that explicitly tracking this many species is not necessary to produce accurate results in certain applications that require non-equilibrium cooling.

Like FLASH, ENZO has been used to study the effects of cooling in the ICM \citep{Li15}.
ENZO is different from the other codes in that it also includes cooling from deuterium, and molecular forms of hydrogen and deuterium; this makes it more suitable for low temperatures below 10\textsuperscript{3} K.
The code can track all species of hydrogen, deuterium, and helium (including the aforementioned molecules).
There is also an option to incorporate the use of look-up tables, which were calculated with the photoionization code CLOUDY \citep{Ferland98}, for metal cooling.
The limitation is that the ENZO code assumes ionization equilibrium for the metals and does not explicitly track number densities like the other codes that track the heavier species.

NIRVANA is another MHD code that has non-equilibrium cooling capabilities.
The implementation is different from other codes in that it reads in a single text file which contains the desired reaction equations.
Cooling is then predicted for each reaction with rate coefficients, which are derived from collision strengths found in various references.
Reaction equations could be implemented to include metals and charge exchange, but the examples given in \citet{Ziegler16} only used the species listed in Table~\ref{tab:codes}.
This implementation method makes it relatively simple to add new species (including molecules), but it becomes computationally expensive when incorporating many species due to the increased number of equations to be solved.
This is a common issue with all hydrodynamic codes: tracking a large chemical network of ionic species becomes computationally expensive, especially when the code is three-dimensional and includes other multi-physical processes.
For this reason, AstroBEAR, and codes similar to AstroBEAR, often use look-up tables to obtain cooling rates.

Not all cooling tables work when the gas is out of ionization equilibrium.
For example, mostly neutral gas at 30000 K cools much faster than ionized gas at the same temperature and density.
For this reason, it is important that cooling routines follow the ionization fraction of H explicitly to give accurate results.

RAMSES is yet another MHD code that, like AstroBEAR, uses AMR \citep{Teyssier02}.
It is an N-body and hydrodynamics code designed to study structure formation at cosmological and galactic scales.
The non-equilibrium cooling implementation is detailed in \citet{Rosdahl13} where the authors have named the extension of this code RAMSES-RT.
Metal cooling is not included in this code in part because it is less important at these large scales.
There are many other codes currently in use, and there are often different versions of a code which are used for different applications.
The main point is that many of these codes treat non-equilibrium cooling in a similar fashion, but they are different in the specific processes and species that they include.
The code user should pay careful attention to what is included in the cooling of a code to determine if it is appropriate for their application.

Section~\ref{sec:ch1methods} contains our numerical methods including the ionization, recombination, non-equilibrium cooling, and production of synthetic emission maps in AstroBEAR.
The best way to assess how well this implementation works is to see how closely it matches the predictions of temperature, density, and magnetic field throughout the post-shock cooling zones from a well-tested 1-D model that includes all the relevant physics.
These comparison models were generated by the shock code built by Cox and Raymond (hereafter referred to as CRSC, \citet{Raymond79}), and they are shown in Section~\ref{sec:1Dshocks}.
We also show some examples of synthetic emission maps from simulations of supersonic outflows in Section~\ref{sec:outflows}.
Finally, in Section~\ref{sec:ch1conc} we summarize our paper and discuss ongoing and future work.

\section{Methods}
\label{sec:ch1methods}
AstroBEAR is a highly parallelized adaptive mesh refinement (AMR) multi-physics code.
The use of AMR is crucial in 2-D and 3-D to resolve regions of rapid cooling and ionization (behind shock fronts).
See \cite{Cunningham09,Carroll12} for a detailed explanation of how AMR is implemented.
More details of the code can also be found at https://astrobear.pas.rochester.edu/trac/.
The code can solve the 3-D MHD equations with a variety of multi-physics and source terms, but for the present work, we will only consider the 3-D ideal MHD equations with non-equilibrium cooling:

\begin{subequations}\label{group1.1}
\begin{gather}
    \frac{\partial \rho}{\partial t} + \boldsymbol{\nabla} \cdot \rho \boldsymbol{v} = 0 ,\ \label{1.1a}\\[\jot]
    \frac{\partial \rho \boldsymbol{v}}{\partial t} + \boldsymbol{\nabla} \cdot (\rho \boldsymbol{v} \boldsymbol{v} + P\boldsymbol{I} - \boldsymbol{B}\boldsymbol{B}) = 0 ,\ \label{1.1b}\\[\jot]
    \frac{\partial E}{\partial t} + \boldsymbol{\nabla} \cdot ((E + P) \boldsymbol{v} - (\boldsymbol{v} \cdot \boldsymbol{B})\boldsymbol{B}) = -L ,\ \label{1.1c}\\[\jot]
    \frac{\partial \boldsymbol{B}}{\partial t} + \boldsymbol{\nabla} \cdot (\boldsymbol{v} \boldsymbol{B} - \boldsymbol{B} \boldsymbol{v}) = 0 ,\ \label{1.1d}\\[\jot]
    \frac{\partial n_i}{\partial t} + \boldsymbol{\nabla} \cdot n_i \boldsymbol{v} = \Gamma_i ,\ \label{1.1e}
\end{gather}
\end{subequations}
where $\rho$ is the mass density, $\boldsymbol{v}$ is the velocity, $P$ is the total pressure (thermal + magnetic) defined as $P = P_{th} + \frac{1}{2} B^2$, $\boldsymbol{I}$ is the identity matrix, $\boldsymbol{B}$ is the magnetic field normalized by $\sqrt{4 \pi}$, and $E$ is the total energy such that $E = \frac{1}{\gamma - 1} P_{th} + \frac{1}{2}\rho v^2 + \frac{1}{2} B^2$ (with $\gamma = \frac{5}{3}$ for an ideal gas).
$L$ is the cooling source term which will be a function of number density, temperature, and ionization.
$n_i$ is the number density of species $i$, and $\Gamma_i$ is the sum of the ionization and recombination rates for species $i$.

The equations above represent the conservation of mass \eqref{1.1a}, momentum \eqref{1.1b}, energy \eqref{1.1c}, and magnetic flux \eqref{1.1d}.
Equation \eqref{1.1e} represents the evolution of the number densities of the different atomic species tracked within the code.
In the subsections that follow, we will describe in detail the recombination and ionization $\Gamma_i$, the cooling source term $L$, and how the inclusion of various species $n_i$ are used to produce synthetic emission maps.

\subsection{Ionization and Recombination}
\label{subsec:ionrecomb}
There currently are a total of 8 species in AstroBEAR whose number densities are tracked via equation \eqref{1.1e}: \HI, \HII, \HeI, \HeII, \HeIII, \SII, \SIII, and \SIV.
The source term $\Gamma_i$ contains the ionization and recombination rates relevant to each species.
Thus, $\Gamma_i$ can be expressed as 8 more equations:

\begin{subequations}\label{group1.2}
\begin{gather}
    \Gamma_{\HI} = -n_e n_{\HI} C_{\HI} + n_e n_{\HII} \alpha_{\HII} ,\ \label{1.2a}\\[\jot]
    \Gamma_{\HII} = n_e n_{\HI} C_{\HI} - n_e n_{\HII} \alpha_{\HII} ,\ \label{1.2b}\\[\jot]
    \Gamma_{\HeI} = -n_e n_{\HeI} C_{\HeI} + n_e n_{\HeII} \alpha_{\HeII} ,\ \label{1.2c}\\[\jot]
    \Gamma_{\HeII} = n_e n_{\HeI} C_{\HeI} - n_e n_{\HeII} \alpha_{\HeII} - n_e n_{\HeII} C_{\HeII} + n_e n_{\HeIII} \alpha_{\HeIII} ,\ \label{1.2d}\\[\jot]
    \Gamma_{\HeIII} = n_e n_{\HeII} C_{\HeII} - n_e n_{\HeIII} \alpha_{\HeIII} ,\ \label{1.2e}\\[\jot]
    \Gamma_{\SII} = -n_e n_{\SII} C_{\SII} + n_e n_{\SIII} \alpha_{\SIII} ,\ \label{1.2f}\\[\jot]
    \Gamma_{\SIII} =  n_e*n_{\SII} C_{\SII} - n_e n_{\SIII} \alpha_{\SIII} - n_e n_{\SIII} C_{\SIII} + n_e n_{\SIV} \alpha_{\SIV} ,\ \label{1.2g}\\[\jot]
    \Gamma_{\SIV} =  n_e n_{\SIII} C_{\SIII} - n_e n_{\SIV} \alpha_{\SIV} .\ \label{1.2h}
\end{gather}
\end{subequations}
Here $n_e$ is the electron number density, $C$ is the collisional ionization rate in units of cm\textsuperscript{3}s\textsuperscript{-1} and $\alpha$ is the sum of the radiative recombination rate and dielectric recombination rate also in units of cm\textsuperscript{3}s\textsuperscript{-1} ($\alpha = \alpha_r + \alpha_d$).
In these equations and everywhere in this paper, the subscript $i$ following $C$ or $\alpha$ refers to the initial species which interacts with an electron and becomes ionized or recombines.

Following the prescription in \citet{Mazzotta98}, the collisional ionization rate for species $i$ can be written as
\begin{equation}
    C_i = \frac{6.69 \times 10^{7}}{(kT)^{3/2}} \frac{\exp{(-x_i)}}{x_i} F(x_i,\sigma_i) , \ \label{1.3}
\end{equation}
where 
\begin{equation}
    x_i = \frac{I_i}{kT} .\ \label{1.4} 
\end{equation}
Here $I_i$ is the ionization potential for species $i$ (e.g. 13.6 eV to ionize \HI into \HII), $T$ is temperature in K, $k$ is the Boltzmann constant in eV/K, and $F$ is a function which depends on $x_i$ and collisional cross section $\sigma_i$.
The collisional cross sections which determine the value of $F$ were taken from \citet{Arnaud85}.

We followed the fit used by \citet{Verner96} to determine radiative recombination rates.
\begin{equation}
    \alpha_r = A \left[ \sqrt{\frac{T}{T_0}} \left( 1 + \sqrt{\frac{T}{T_0}} \right) ^{1-b} \left( 1 + \sqrt{\frac{T}{T_1}} \right) ^{1+b} \right] ^{-1} ,\ \label{1.5}
\end{equation}
where $A$, $T_0$, $T_1$, and $b$ are fitting parameters different for each species.
Dielectric recombination is different from radiative recombination in that the ion first transitions to an excited state and then to its ground state.
\citep{Mazzotta98} gives the dielectric recombination rate as 
\begin{equation}
    \alpha_d = \frac{c}{T^{3/2}} \exp{\left(-\frac{E}{T}\right)} ,\ \label{1.6}
\end{equation}
where $c$ and $E$ are the fitting parameters.
The sum of $\alpha_r$ and $\alpha_d$ gives the total recombination rates $\alpha$ used in equations \eqref{group1.2}.

It is also important to note how we keep track of the electron number density $n_e$.
The gas is assumed to be quasi-neutral, thus the electron number density $n_e$ is equal to 
\begin{equation}
    n_e = \max(n_{\HII} + n_{\HeII} + 2 n_{\HeIII} , n_{min}) ,\ \label{1.7}
\end{equation}
where $n_{min}$ is a minimum density which we chose to be 0.01 times the total gas density.
This minimum is necessary to ensure that there are always at least some electrons present to ionize the gas.
The use of such a minimum is also physically realistic since the heavier metals (e.g., S or Fe) will always have some level of ionization due to the presence of far-ultraviolet (FUV) photons.
This is also why we do not track neutral \SI; we assume that all sulfur is at least singly ionized (i.e., \SII).

All of the collisional ionization rates $C$ and the recombination rates $\alpha$ multiplied by the appropriate number densities form the right hand sides of equations \eqref{group1.2}.
AstroBEAR first solves the source-free version of equation \eqref{1.1e} (i.e. with $\Gamma = 0$) and then uses an operator split approach to handle the microphysical source terms.
The source terms $\Gamma$ are integrated using a 5\textsuperscript{th} order accurate Cash-Karp Runge-Kutta method.
This method uses the difference between the 4\textsuperscript{th} and 5\textsuperscript{th} order solutions to determine accuracy and adjust the time-step as appropriate.
AstroBEAR uses the same method for other constant source terms (e.g., the cooling term $L$).

\subsection{Non-equilibrium Cooling}
\label{subsec:cooling}
The cooling source term $L$ in equation \eqref{1.1c} is treated in the same way as $\Gamma$ in equation \eqref{1.1e}.
As described in the previous subsection, the source term is integrated using a Cash-Karp Runge-Kutta method.
$L$ is comprised of four main categories of coolants: hydrogen $L_H$, helium $L_{He}$, other heavy metals $L_Z$, and molecules $L_M$ as illustrated in equation \eqref{1.8}.
\begin{equation}
    L = L_H + L_{He} + L_Z + L_M  .\ \label{1.8}
\end{equation}

Each of these coolants can be further separated into more terms via the underlying processes responsible for the energy losses.

$L_H$ contains energy losses due to \HI \ excitation, \HI \ ionization, and \HII \ recombination.
Thus, $L_H$ can be written as
\begin{equation}
    L_H = \HI_{ex} + n_e \left( n_{\HI} C_{\HI} I_{\HI} + n_{\HII} \alpha_{\HII} k T \right)  .\ \label{1.9}
\end{equation}
Likewise, $L_{He}$ can be split into its excitation, ionization, and recombination terms:
\begin{equation}
    L_{He} = \HeI_{ex} + \HeII_{ex} + n_e \left( n_{\HeI} C_{\HeI} I_{\HeI} + n_{\HeII} \alpha_{\HeII} k T + n_{\HeII} C_{\HeII} I_{\HeII} + n_{\HeIII} \alpha_{\HeIII} k T\right) .\ \label{1.10}
\end{equation}
In equations \eqref{1.9} and \eqref{1.10}, the terms labeled by subscripts `$ex$' represent cooling due to excitation.
In previous versions of AstroBEAR, these were taken directly from the Dalgarno \& McCray (DM) cooling curve which assumes ionization equilibrium values for the number densities \citep{Dalgarno72}.
Through the use of equation \eqref{1.1e}, we can now use non-equilibrium values for the number densities which results in more accurate energy losses.

To convert the ionization and recombination rates, $C$ and $\alpha$ respectively, into energy loss rates we multiply by appropriate number densities and energies.
The ionization terms are multiplied by the ionization potential $I_i$ for species $i$, and the recombination terms are multiplied by a factor of $2/3 kT$.
This factor of 2/3 warrants additional explanation.

Although the average kinetic energy of the electrons is $3/2 kT$ and the most probable kinetic energy is $kT$, the recombination cross sections of ions (including H) decline as electron energies increase.
In general, values of the average energy lost per recombination to level $n$ depend both upon $n$ and the temperature, and vary from about $0.9 kT$ to $0.5 kT$ as $n$ changes from 2 to 6 for H at temperatures of interest.
Hence we adopt 2/3 for the multiplicative constant in front of the $kT$ term for recombination losses.
Typically, collisional cooling dominates by a factor of $\sim$ 10 over recombination, so the exact value of the constant has a minor effect on the overall energy budget in the simulations.

Cooling from metal excitation is calculated from one of two different tables depending on the temperature.
In between 2000 and 16500 K, we use a new cooling table which contains forbidden line cooling from \OI, \OII, \NI, \NII, \SII, \FeII, \SiII, \MgII, and \CII.  
The table uses strong charge exchange cross sections to lock the ratios of \NII/\NI \ and \OII/\OI \ to \HII/\HI, and it solves multilevel atom models to derive volume emission cooling terms.
This 3-D table depends on the electron number density, ionization fraction, and temperature.
Hence the tracking of various neutral and ionized species makes the use of this table possible and results in more accurate cooling due to metals.
Cooling in this table (hereafter referred to as ``Z'' cooling) is strictly from metal-electron collisions.

\begin{figure}
\centering
\includegraphics[width=0.75\linewidth]{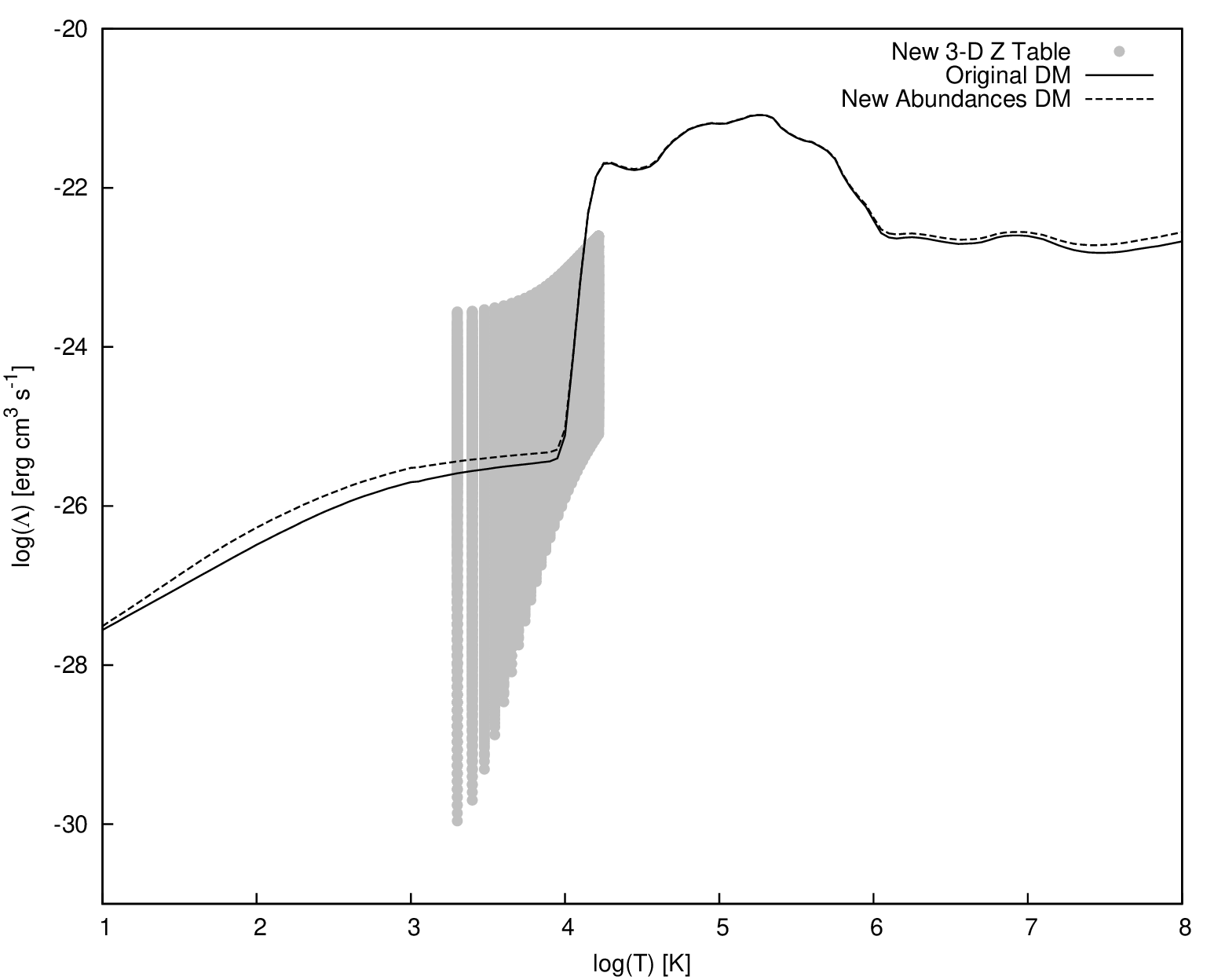}
\caption{Cooling curves used in AstroBEAR.
The y-axis shows the cooling efficiency $\Lambda$ which leads to the energy loss term $L$ when multiplied by appropriate number densities ($n_e$ and total hydrogen number density $n_\text{H}$).
The \emph{solid} line represents the original DM cooling curve used in previous versions of AstroBEAR, \emph{dashed} line is the DM curve but with the slightly altered abundances, and the \emph{gray circles} are all of the data points from the new 3-D metal cooling table.
The upper boundary of the gray area represents the fully ionized limit, while the lower boundary represents the fully neutral limit.
The new table allows for the possibility of having gas far from ionization equilibrium in the region where most of the forbidden lines originate and where magnetic pressure effects are greatest.}
\label{fig:coolingcurves}
\end{figure}

We also use a modified version of the Dalgarno \& McCray (DM) cooling curve \citep{Dalgarno72} where we subtract the contributions from H and He, leaving only the metal and molecular components.
We altered the abundances to match the abundances used in the aforementioned 3-D metal cooling table which are more appropriate for the interstellar medium (see Table~\ref{tab:abundances}).
The metal and molecular components of the DM curve are further altered by using non-equilibrium densities instead of the values one would get from ionization equilibrium.
At every temperature, we use this modified DM curve for metal-hydrogen collisional cooling and molecular cooling, although the molecular component will only become important below 1000 K.
Above 16500 K and below 2000 K, the metal-electron collision term of the modified DM curve is used instead of the value from the 3-D table.
A smoothing function is applied at the boundaries of these two metal-electron cooling tables (2000 and 16500 K) to avoid discontinuous jumps in cooling rates.
Figure~\ref{fig:coolingcurves} shows the following cooling curves: DM, an altered DM, and the new 3-D Z cooling table.

\begin{deluxetable}{cc}
    \tablenum{2}
    \tablecolumns{2}
    \tablewidth{0pt}
    \tablecaption{Element abundances relative to hydrogen \label{tab:abundances}}
    \tablehead{\colhead{Element} & \colhead{Abundance}}
    \startdata
        H & 12 \\
        C & 8.52 \\
        N & 7.96 \\
        O & 8.82 \\
        Ne & 2.60 \\
        Mg & 7.42 \\
        Si & 7.52 \\
        S & 7.20 \\
        Fe & 7.60
    \enddata
\end{deluxetable}

\subsection{Calculating Synthetic Emission Lines}
\label{subsec:emiss}
In the following subsection, all emission equations give an emissivity in units of erg/cm\textsuperscript{3}/s.
Tracking the hydrogen and helium species allows us to track their ionization fractions and thus the electron number density of the gas which is required for generating synthetic emission maps (equation \eqref{1.7}).
For HH objects, [\SII] emission is strongest in cooling regions behind shocks \citep{Heathcote96}.
In AstroBEAR, below $10^4 K$, all S is assumed to be \SII, and above $10^4 K$, the ionization and recombination rates are used to track the amount of \SIII \ and \SIV.
This was employed to more accurately track the ionization state of S at higher temperatures and hence produce more accurate [SII] maps.

Our total [\SII] emission is the sum of two lines with photon wavelengths of 673.1 and 671.6 nm.
These correspond to electron transitions from level 2 to level 1 and level 3 to level 1, respectively.
The total [\SII] emission can be written as
\begin{equation}
    [\text{\SII}] = \sum_{j=2}^{3} n_j A_j h \nu_j ,\ \label{1.11} 
\end{equation}
where $n_j$ is the number density of \SII \ in excited state $j$, $A_j$ is the Einstein A-coefficient for spontaneous emission from level $j$ to level 1, $h$ is Planck's constant, and $\nu_j$ is the frequency of the emitted photon such that $h\nu_j$ is in ergs.
The values for $A_j$ are taken from \citet{Mendoza82}, and we solve the 5-level atom to determine the values for $n_j$.
In solving for the level populations $n_j$ of this system, we used collision strengths from \citet{Keenan96}.

Calculating the H$\alpha$ emission accurately is crucial to comparing these simulations with observations.
H$\alpha$ will typically mark shock fronts within HH objects \citep{Heathcote96}.
In AstroBEAR, the H$\alpha$ routine is dependent on electron number density, temperature, ionization fraction, and total hydrogen number density.
The total H$\alpha$ emission is the sum of two terms: a collisional excitation term H$\alpha_c$ and a recombination term H$\alpha_r$.

For the excitation term, we include excitations to levels 3, 4, and 5 using effective collision strengths $\upsilon$ from \citet{Anderson00} and \citet{Anderson02}.
The excited atom then de-excites, and an H$\alpha$ photon is produced when the atom transitions from level 3 to level 2.
Such a photon has energy $h\nu \simeq 3.0263 \times 10^{-12}$ ergs.
The electrons in the higher levels (4 and 5) have to cascade down to level 3 and then level 2 to be included in the H$\alpha$ emission, thus their contributions are less significant than electrons already in level 3.
The excitation term can be written as
\begin{equation}
    \text{H}\alpha_c = n_e n_{\HI} h \nu \frac{8.63 \times 10^{-6}}{2\sqrt{T}} \sum_{j=3}^{5} \exp{\left(\frac{-E_j}{kT}\right)}\upsilon_j ,\ \label{1.12}
\end{equation}
where $E_j$ is the energy required to excite neutral H from level 1 to level $j$, and $\upsilon_j$ is the collision strength for such an excitation.
The collision strength includes a factor which accounts for the fact that not all excitations lead to H$\alpha$ emission; some will produce H$\beta$ and H$\gamma$ emission as well.

The recombination term is important when the gas is highly ionized at relatively low temperatures $\lesssim$ $10^4$ K.
When ionized hydrogen is recombining, there is a nonzero probability that it will go through the H$\alpha$ transition.
To calculate this term, the radiative recombination rate $\alpha_r$ is simply multiplied by the appropriate quantities as follows:
\begin{equation}
    \text{H}\alpha_r = n_e n_{\HII} h \nu \alpha_r .\ \label{1.13}
\end{equation}
As in Subsection~\ref{subsec:ionrecomb}, recombination rates are taken from \citet{Verner96}, and they are valid above 3000 K.

Both recombination and collisional excitation can be important depending on the conditions.
In general, at low temperatures ($\lesssim$ $10^4$ K), H$\alpha$ emission is weak but is dominated by the recombination term because any hydrogen that is ionized will quickly recombine.
The excitation term dominates at higher temperatures (up to 10\textsuperscript{5} K) as the gas is less likely to recombine and energies are favorable for excitation.
At still higher temperatures, the amount of H$\alpha$ will decrease because the hydrogen becomes mostly ionized (no neutral H to excite).

AstroBEAR does all of the emission calculations as post-processing after each outputted frame from a simulation.
In order to generate a synthetic emission map, the gas is assumed to be optically thin, so the emission can be easily summed along a line of sight.
The data can be rotated or inclined to produce a 2-D emission map from different observer perspectives.
The emission maps in this paper show [\SII] in red and H$\alpha$ in green, and the ratio [\SII]/H$\alpha$ determines the color such that when this ratio approaches unity, the color will become yellow.

\section{Testing AstroBEAR's Cooling Routines}
\label{sec:1Dshocks}
One of the first problems that one might study with an MHD code that uses equations~\eqref{group1.1} is a 1-D, stationary, magnetized, radiative shock.
Exploring the behavior of such shocks with varying initial conditions helps us understand the various environments where these shocks exist such as within YSO jets.
However, to make comparisons with observations, it is necessary for a code to accurately track temperature, densities, and other quantities in the post-shock region where the gas is cooling.
We ran a number of models with AstroBEAR and compared the results with equivalent models run by CRSC.

For our AstroBEAR models, we let the gas move in the x-direction with velocity $v_x$ and the magnetic field is in the y-direction $B_y$.
This was chosen because any field component that is perpendicular to the shock front ($B_x$) will not be compressed by the shock.
Given initial pre-shock conditions for density, velocity, temperature, field strength, and ionization fraction, the 1-D shock jump equations can be used to find the post-shock values.
The 1-D MHD shock jump equations are as follows:
\begin{subequations}\label{group1.14}
\begin{gather}
    \left[ \rho v_x \right]^2_1= 0 ,\ \label{1.14a}\\[\jot]
    \left[ \rho v_x^{\ 2} + P_{th} + \frac{B_y^{\ 2}}{2} \right]^2_1 = 0 ,\ \label{1.14b}\\[\jot]
    \left[ (\frac{1}{2}\rho v_x^{\ 2} + \frac{\gamma}{\gamma - 1}P_{th} + B_y^{\ 2}) v_x \right]^2_1 = 0 ,\ \label{1.14c}\\[\jot]
    \left[ B_y v_x \right]^2_1 = 0 ,\ \label{1.14d}\\[\jot]
    \left[ n_i v_x \right]^2_1 = 0 ,\ \label{1.14e}
\end{gather}
\end{subequations}
where the variables are defined in the same way as they were in Section~\ref{sec:ch1methods} for equations~\eqref{group1.1}.
The bracket notation is used to define the difference between post-shock and pre-shock quantities ($\left[ A \right]^2_1 = A_2 - A_1$).
For simplicity, the ionization fraction is assumed to remain constant across the shock jump.

The cooling and ionization/recombination source terms must be accounted for in the post-shock region.
We use a Runge-Kutta integration method to solve for the energy flux and species number density fluxes one computational cell at a time.
From these values, we can derive a density, velocity, temperature, and magnetic field to populate all of the cells behind the shock.
This can be done out to any distance, but we stop the calculation once the temperature drops to about 1000 K since this is where molecular cooling processes would start to become important.
We will refer to the distance from the shock front to T = 4000 K as the cooling length $L_{cool}$.
Once the domain is initialized with the pre-shock region, shock jump, and post-shock region, the code can begin taking hydrodynamic time-steps.
Since the simulations are run in the frame of the shock, the entire radiative shock profile remains stationary and does not change.

The CRSC \citep{Raymond79} treatment of a radiative shock is very different than that of AstroBEAR.
It is not an MHD code; it instead follows a single parcel of gas as it travels through a post-shock region and radiates.
CRSC, as compared to AstroBEAR, contains many more emission lines, explicitly tracks more ionized species, and accounts for more chemical processes such as photoionization.
The CRSC also calculates non-local radiative transfer by following the optical depth of Lyman continuum photons.
Such photons are mostly produced at the shock front and can reionize the gas in the post-shock cooling zone.
This effect is mitigated in AstroBEAR by limiting the shock velocities to below 90 km/s.
Modeling higher shock velocities would require a treatment of photoionization and radiative transfer in 3-D which can be computationally expensive.
Higher shock velocities also require higher resolution which imposes a code limitation to which shock models we can reasonably compare.
Despite these differences and limitations, radiative shock models of modest shock velocities ($<$ 90 km/s) from both codes are strikingly similar.
Furthermore, the shock velocity limitation still allows for accurate modeling of HH objects.
Although HH objects exhibit speeds of 200-400 km/s (determined by proper motions and radial velocities), the knots within such objects move into previously ejected material with relative velocities on the order of 10-100 km/s \citep{Hartigan11}.

We tested the new AstroBEAR cooling routine by running dozens of models, and highlight four of them here which vary in pre-shock ionization fraction and shock velocity (see Table~\ref{tab:shockmodels}).
Changing the ionization fraction is the best way to test the ionization/recombination routines in different regimes in AstroBEAR, and changing the shock velocity is the best way to test different temperature ranges.
All four models have the same pre-shock values for total atomic number density and temperature of 100 cm\textsuperscript{-3} and 10000 K, respectively. 
Models A and B have shock velocities of 40 km/s and models C and D have shock velocities of 70 km/s.

\begin{deluxetable}{cccc}
    \tablenum{1.2}
    \tablecolumns{4}
    \tablewidth{0pt}
    \tablecaption{Varied parameters in 1-D shock models \label{tab:shockmodels}}
    \tablehead{\colhead{Model} & \colhead{v [km/s]} & \colhead{B [$\mu$G]} & \colhead{X}}
    \startdata
        A & 40 & 94 & 0.01 \\
        B & 40 & 94 & 0.40 \\
        C & 70 & 77 & 0.01 \\
        D & 70 & 77 & 0.40 \\
    \enddata
\end{deluxetable}

\begin{figure}
\centering
\includegraphics[width=0.7\linewidth]{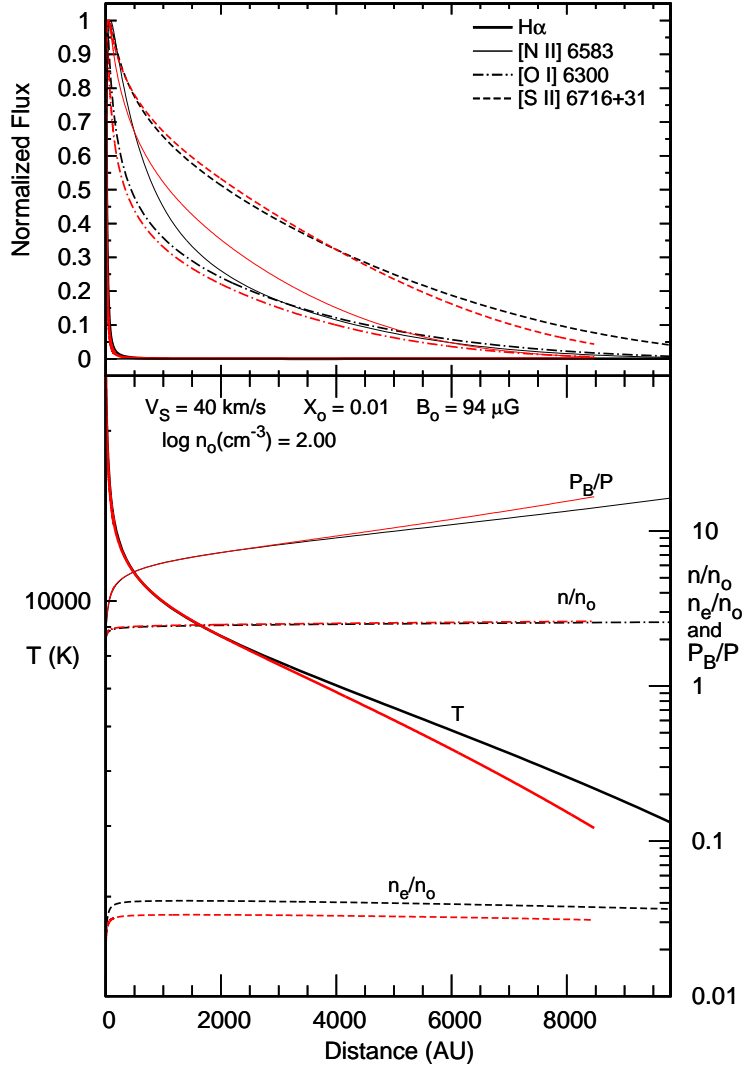}
\caption{Radiative shock profile of model A.
The \emph{red} lines are from the AstroBEAR model, and the \emph{black} lines are from CRSC.
The top plot shows the emission lines, and the bottom plot shows the temperature $T$, number density compression ratio $n/n_o$, inverse plasma beta $P_B/P$, and electron number density ratio $n_e/n_o$.
The temperature values use the left vertical axis, and the ratios use the right vertical axis.
Comparing $n/n_o$ to $n_e/n_o$ gives a sense for how ionization fraction changes with distance.
The inverse plasma beta $P_B/P$ is greater than 1 which means that the magnetic field pressure dominates over thermal pressure forces.}
\label{fig:lowXlowM}
\end{figure}

Figure~\ref{fig:lowXlowM} shows the radiative shock profile for model A.
This figure, and the three that follow, show the post-shock profiles for several MHD quantities and five different emission lines.
We have plotted the temperature in K, density compression $n/n_o$, ionization fraction $n_e/n_o$, and inverse plasma beta $P_B/P$.
As expected, the temperature drops and the density increases, hence the increase in thermal pressure is minimal.
The magnetic pressure, however, increases along with the density due to magnetic flux freezing \citep{Hartigan15}, thus $P_B/P$ increases in the post-shock region.

The emission lines that we have plotted here include H$\alpha$, [\OI] (630.0 nm), [\NII] (658.3 nm), and [\SII] (671.6 nm + 673.1 nm).
H$\alpha$ always peaks at the shock front, as expected, and [\SII] peaks farther from the shock front.
Within the post-shock cooling region, the [\SII] emission is much stronger than H$\alpha$.
Oxygen and nitrogen tracers have not been implemented in AstroBEAR as was done for sulfur, so the [\OI] and [\NII] lines are considered to be upper limits.
Even without additional tracers, the code predicts [\OI] and [\NII] emission reasonably well for some models as compared to the CRSC values.

\begin{figure}
\centering
\includegraphics[width=0.7\linewidth]{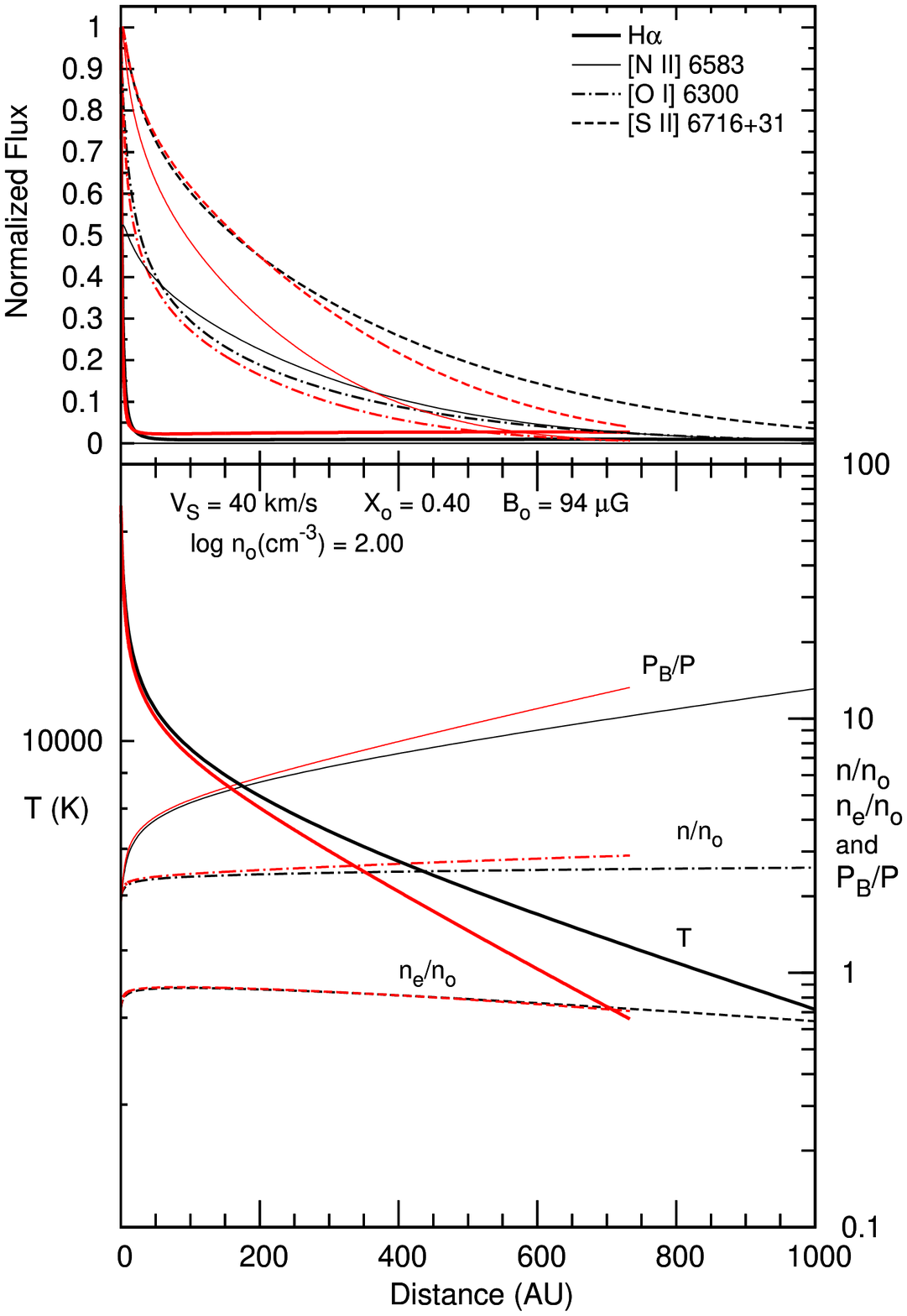}
\caption{Radiative shock profile of model B.
The \emph{red} lines are from the AstroBEAR model, and the \emph{black} lines are from CRSC.}
\label{fig:highXlowM}
\end{figure}

Model B (Figure~\ref{fig:highXlowM}) shows how a higher initial ionization fraction affects the shock jump and post-shock quantities.
Higher ionization leads to a lower initial post-shock temperature which is counter-intuitive but a correct result of how we chose our models.
One would expect higher ionization to lead to higher temperatures, which is true in most of the post-shock cooling zone and is evidenced by the emission lines \citep{Cox85}.
The differences in the initial post-shock temperatures in our models can be explained via the mean atomic weight.
The presence of more electrons decreases the mean atomic weight which in turn increases the sound speed.
Since the shock velocity and pre-shock temperature are fixed, the shock for Model B is effectively weaker than Model A (lower Mach number), hence the temperature jump is less.
For a more straightforward comparison, the initial post-shock temperatures are listed in Table~\ref{tab:Lcools}.
In the post-shock region of Model B, the ionization fraction increases initially as the gas equilibrates to the high temperature, and then it begins to decrease as recombination begins to dominate.
Also note that with this higher ionization, the metal emission lines emit over a longer extent in the post-shock cooling region.

For Model C, we again have a low ionization fraction as in Model A, but we have increased the shock velocity to 70 km/s.
The higher shock velocity means that this shock is stronger, thus the temperature and density jumps are higher.
These initial post-shock temperatures are so high that the ionization processes are dominant, and the gas quickly reaches ionization fractions that are even higher than Model B.
We again see that despite the higher shock velocity and temperatures, AstroBEAR agrees fairly well with the CRSC models in all MHD quantities.
The emission line profiles are also in good agreement, with the H$\alpha$ and [\SII] lines being most accurate.

\begin{figure}
\centering
\includegraphics[width=0.7\linewidth]{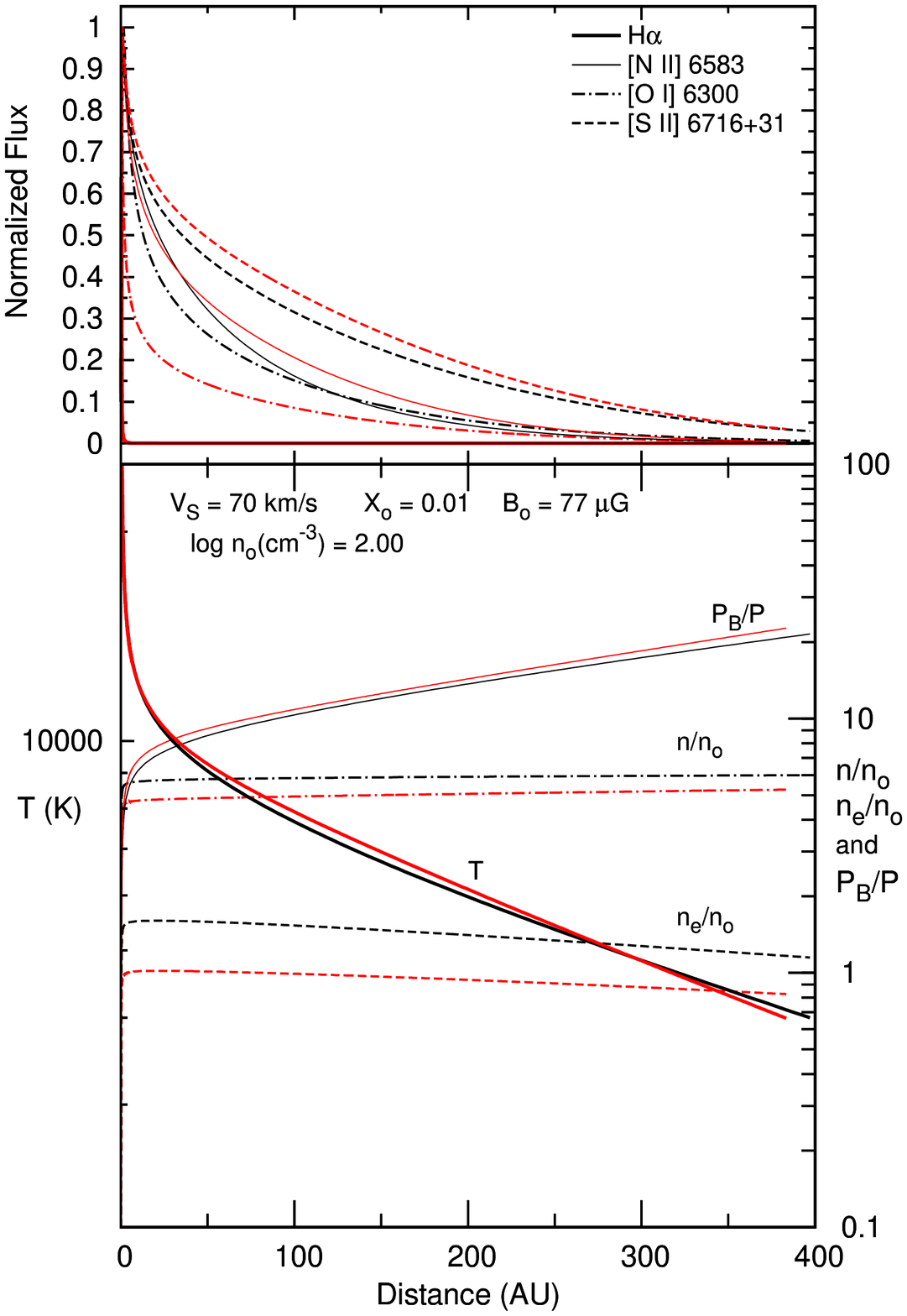}
\caption{Radiative shock profile of model C.
The \emph{red} lines are from the AstroBEAR model, and the \emph{black} lines are from CRSC.}
\label{fig:lowXhighM}
\end{figure}

\begin{figure}
\centering
\includegraphics[width=0.7\linewidth]{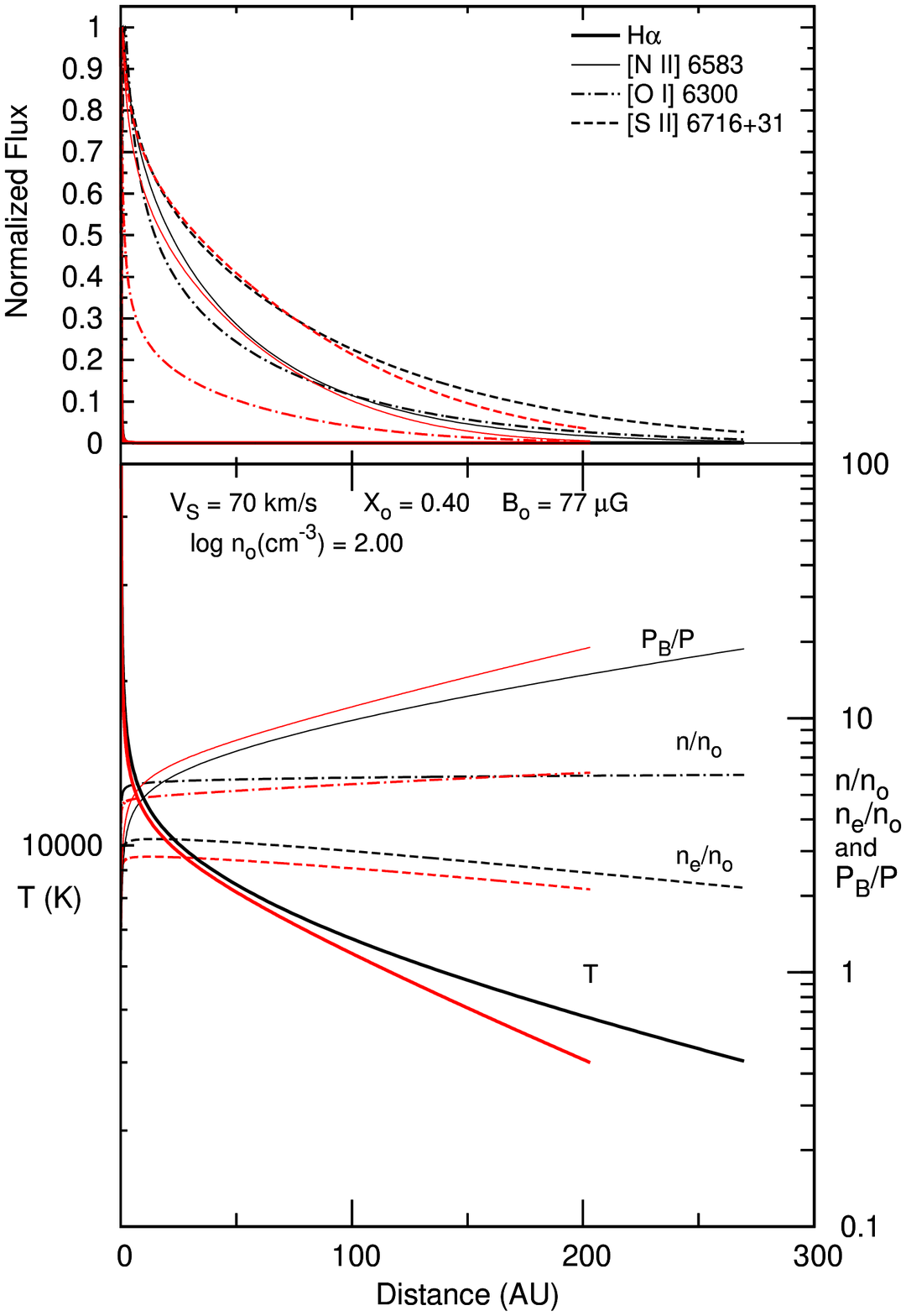}
\caption{Radiative shock profile of model D.
The \emph{red} lines are from the AstroBEAR model, and the \emph{black} lines are from CRSC.}
\label{fig:highXhighM}
\end{figure}

Figure~\ref{fig:highXhighM} shows Model D, the high ionization, high shock velocity model.
As we saw before, the higher ionization actually decreases the temperature and density jump.
This particular model is the most extreme case, and AstroBEAR still captures the correct trends.
Higher shock velocity and higher ionization models would be more challenging for AstroBEAR as \OIII~ and \NIII~ start to become important in these regimes.
In the future, tracers could be added to more accurately model velocities $\geq 90$ km/s and pre-shock ionization fractions $\geq 0.5$.

An important result from running these models is obtaining a value for the cooling length $L_{cool}$
We typically define $L_{cool}$ as the distance from the shock front at which the gas has cooled to a certain temperature, and here we have defined the cooling length temperature to be 4000 K.
Cooling lengths were calculated for all four models for both AstroBEAR and CRSC, and the results are reported in Table~\ref{tab:Lcools}.
On average, the values for $L_{cool}$ from both codes vary by approximately 20\% relative error.
AstroBEAR does appear to cool faster, but it still recovers the expected behaviors and trends of radiative shocks.

\begin{deluxetable}{cc p{40pt}|cc}
    \tablenum{1.3}
    \tablecolumns{5}
    \tablewidth{0pt}
    \tablecaption{Cooling lengths and initial post-shock temperatures of the 1-D shock models \label{tab:Lcools}}
    \tablehead{& \multicolumn{2}{c}{Cooling length (AU)} & \multicolumn{2}{c}{Post-shock temperature (10\textsuperscript{4} K)} \\
    \colhead{Model} & \colhead{AstroBEAR} & \colhead{CRSC} & \colhead{AstroBEAR} & \colhead{CRSC}}
    \startdata
        A & 8417 & 9793 & 2.54 & 2.50 \\
        B & 732 & 1029 & 2.18 & 2.12 \\
        C & 383 & 397 & 11.32 & 11.24 \\
        D & 203 & 270 & 8.60 & 8.55 \\
    \enddata
\end{deluxetable}

For any simulation that uses cooling, it is important to know the cooling lengths of the strongest radiative shocks within the grid because stronger shocks cool faster and will have a shorter $L_{cool}$.
As discussed in \citet{Hansen15b}, the cooling length resolution is crucial to resolving instabilities caused directly or indirectly by strong cooling.
\citet{Yirak10} found that a cooling length resolution of at least 10 cells per $L_{cool}$ was required to resolve such instabilities.
For any AstroBEAR simulation that uses the new non-equilibrium cooling functions, a separate 1-D radiative shock simulation seeded with parameters from the full 2-D or 3-D simulation of interest can be a helpful guide.
The 1-D models can give a value for $L_{cool}$ and thus a minimum for what the physical resolution should be if we want to resolve cooling processes within post-shock regions which will also lead to better values for the emission line calculations.

\begin{figure}
\centering
\includegraphics[width=0.75\linewidth]{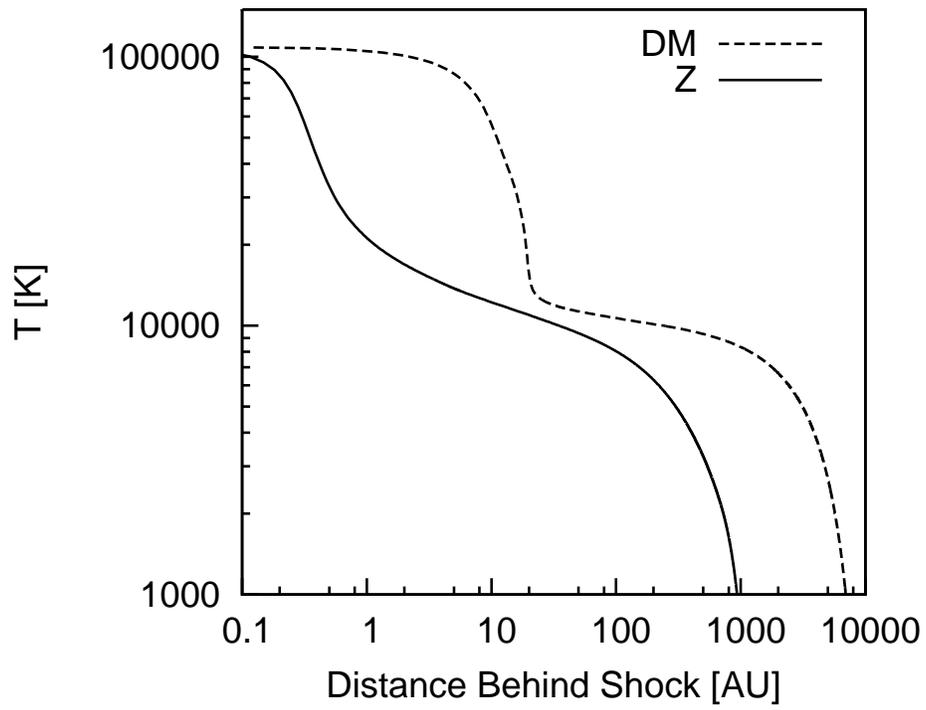}
\caption{Improvement from previous DM cooling in AstroBEAR.
The shock profile shown for comparison here is model C.
The \emph{solid} line is the new non-equilibrium cooling in AstroBEAR, the \emph{dashed} is from the previously used DM cooling.
Z cooling appears to be much stronger and leads to a much shorter cooling length for this shock.}
\label{fig:DMcompare}
\end{figure}

In Figure~\ref{fig:DMcompare} we compare a shock profile from the new non-equilibrium cooling to the previously used DM cooling.
This Figure only shows temperature down to 1000 K, and it is for model C.
We see that with the new implementation, AstroBEAR cools somewhat faster, especially in the temperature range of the new 3-D metal cooling table.
The new implementation is also more consistent with CRSC; the DM cooling model resulted in a cooling length that is approximately 10 times too long.
DM cooling is still quite good at high temperatures, but it is clear that a non-equilibrium treatment is necessary for radiative shocks.
One of the main reason non-equilibrium cooling is important for this application is that neutral H and He cool rapidly at high temperatures.

\section{Applications to HH Objects}
\label{sec:outflows}
An important application of the new non-equilibrium cooling routines in AstroBEAR is simulations of HH objects.
In previous studies, we have run simulations of pulsed jets \citep{Hansen15b} and interacting bow shocks \citep{Hansen17}.
The example outflow and synthetic emission map we show here is similar to the simulations from the latter work.

We initialize a 3-D grid with an overdense clump of radius 10 AU which moves supersonically, at M = 15, producing a radiative bow shock.
The stationary ambient number density, temperature, and ionization fraction are 1000 cm\textsuperscript{-3}, 1250 K, and 0.01 respectively.
These values are consistent with typical values in regions with YSO jets \citep[see][and references therein]{Frank14}.
The clump is initialized with a density of 5 x 10\textsuperscript{5} cm\textsuperscript{-3}, temperature of 2000 K, and ionization fraction of 10\textsuperscript{-3}.
The choice of initial conditions here is not crucial since the purpose of the simulations was to drive a radiative bow shock into the ambient.
Supersonic clumps are a simple way to generate bow shocks in simulations, and these values do not affect the conclusions we can make about the emission from such bow shocks.

\begin{figure}
\centering
\includegraphics[width=0.5\linewidth]{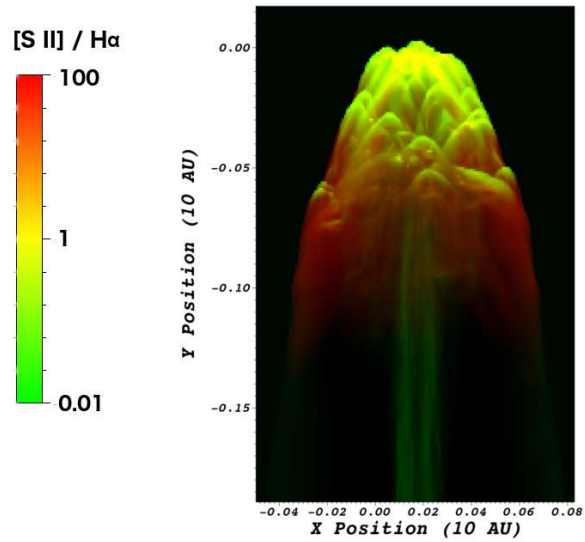}
\caption{Synthetic emission map from a radiative bow shock.
The ratio [\SII]/H$\alpha$ is plotted such that green shows H$\alpha$, red shows [\SII], and yellow shows regions where this ratio approaches unity.
The bow shock was created in simulation by launching a supersonic, over-dense clump.}
\label{fig:emissmap}
\end{figure}

We see the emission from a radiative bow shock in Figure~\ref{fig:emissmap} where H$\alpha$ is in green and [\SII] is in red.
The emission map plots the ratio [\SII]/H$\alpha$ as a color such that it is yellow when the ratio approaches unity.
We see that the green H$\alpha$ is strongest near the shock front, and the emission reddens as the gas cools in the post-shock region as expected.
This is consistent with the 1-D radiative shock models from the previous Subsection.

\begin{figure}
\centering
\includegraphics[width=0.5\linewidth]{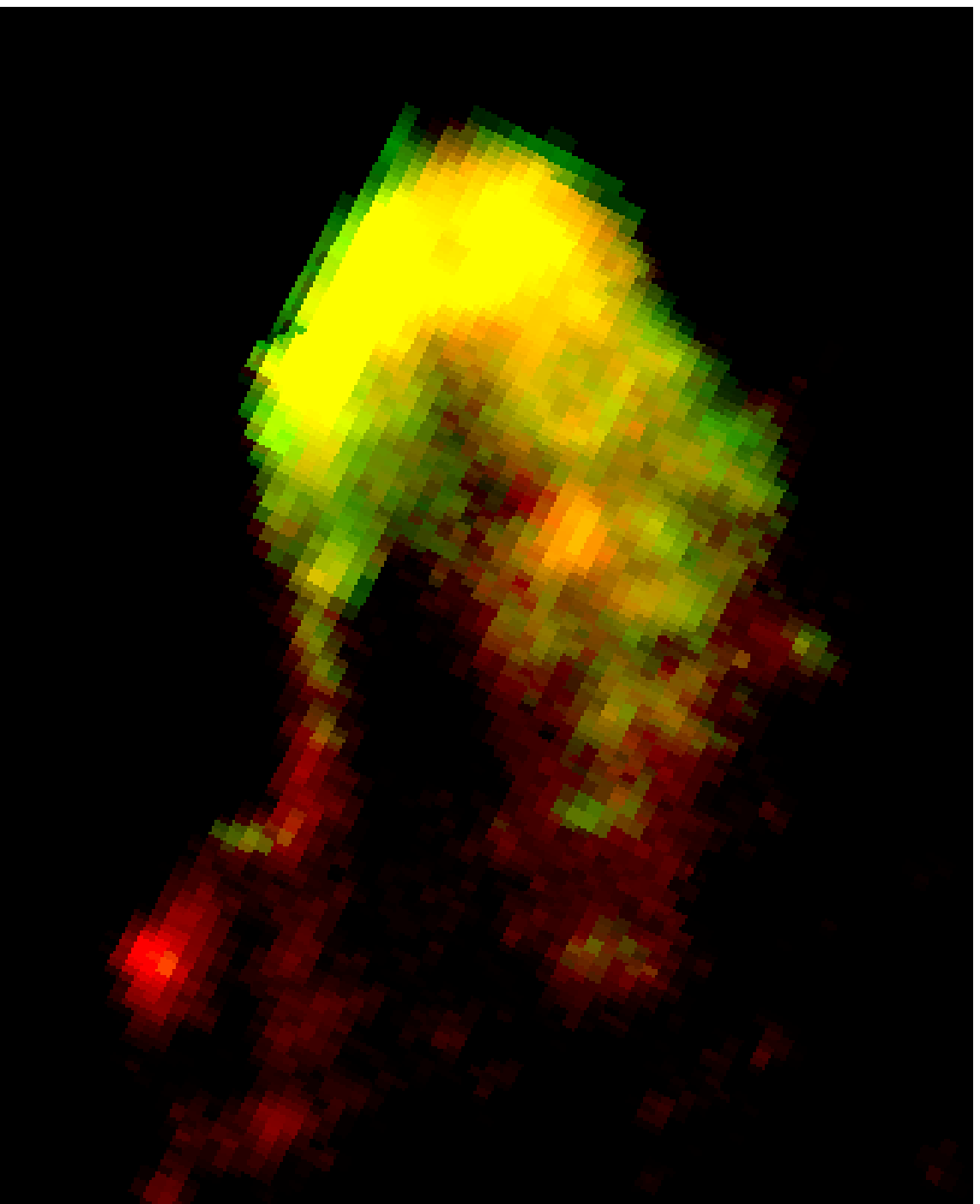}
\caption{HST image of HH 1 from 2007.
This bow shock shows H$\alpha$ and [\SII] features that are similar to those of the synthetic emission map of Figure~\ref{fig:emissmap}.}
\label{fig:HHcompare}
\end{figure}

The use of non-equilibrium cooling and synthetic emission map production make it possible to compare simulation results with actual HST observations of HH objects.
Figure~\ref{fig:HHcompare} shows an HST image of the bow shock of HH 1 from 2007.
We again see bright H$\alpha$ at the shock front and [\SII] dominates in the post-shock cooling region.
There are other similarities between Figure~\ref{fig:HHcompare} and Figure~\ref{fig:emissmap} such as the heterogeneous nature of the emission.
These structures and other details of such outflows will are discussed in \citep{Hansen17}.

\section{Conclusions}
\label{sec:ch1conc}
We have presented the implementation of new non-equilibrium cooling routines into the MHD code AstroBEAR.
These routines are robust and include ionization and recombination equations for various atomic species.
The accuracy of the code was tested by comparing 1-D radiative shock models with CRSC.
Despite large differences in the way each code handles the physics, the models were very similar with an average relative differences in cooling length of approximately 18\%.
1-D radiative shock models can also be used as a tool to determine the cooling length in other simulations.
It is important to resolve $L_{cool}$ in order to accurately simulate cooling processes, including thermal instabilities, within post-shock regions.

We also presented the implementation of post-processing routines in AstroBEAR built to produce synthetic emission maps of H$\alpha$ and [\SII].
The [\OI] and [\NII] lines are implemented in the same way as [\SII], and more lines may be added in the future.
To produce other emission line fluxes, tracers for other species (e.g. \OIII) would have to be implemented, and the framework for implementing tracers already exists in the current version of AstroBEAR.
With an accurate measure of temperature and ionization in post-shock regions, synthetic emission maps from AstroBEAR can be directly compared to HST images.

We are currently using the new non-equilibrium cooling functionality in other research projects such as 3-D pulsed jets which will expand on the work done in \citet{Hansen15b}.
There is much research that can be done with the current implementation, but further improvements to cooling in AstroBEAR can still be made.
The cooling routines can be expanded to include more processes and track more species within the temperature regions relevant in this paper.
Furthermore, much work is needed to implement accurate molecular cooling if simulations are to be conducted at temperatures below 1000 K.
Molecular cooling is a challenging problem as it would need to include processes such as grain formation and photodissociation.

\vspace{7 mm}
\noindent\emph{Acknowledgements.} This work was supported by the Center for Integrated Research and Computing at the University of Rochester which provided computational resources.
Financial support for this project was provided by the Laboratory for Laser Energetics at the University of Rochester; by Space Telescope Science Institute grants HST-AR-11251.01-A (2007), HST-AR-12128.01-A, and HST-AR013892.001-A; by the National Science Foundation under award AST-0807363; by the Department of Energy National Laser User Facility (NLUF) grants DE-NA0002037 and DE-NA0002722; by the Department of Energy under award DE-SC0001063.

\bibliographystyle{apj}
\bibliography{mylibrary}

\begin{thebibliography}{33}
\expandafter\ifx\csname natexlab\endcsname\relax\def\natexlab#1{#1}\fi

\bibitem[{Anderson {et~al.}(2000)Anderson, Ballance, Badnell, \&
  Summers}]{Anderson00}
Anderson, H., Ballance, C.~P., Badnell, N.~R., \& Summers, H.~P. 2000, JPhB,
  33, 1255A

\bibitem[{Anderson {et~al.}(2002)Anderson, Ballance, Badnell, \&
  Summers}]{Anderson02}
---. 2002, JPhB, 35, 1613A

\bibitem[{Arnaud \& Rothenflug(1985)}]{Arnaud85}
Arnaud, M. \& Rothenflug, R. 1985, A\&A, 60, 425

\bibitem[{Bryan {et~al.}(2014)Bryan, Norman, O'Shea, Abel, Wise, Turk,
  Reynolds, Collins, Wang, Skillman, Smith, Harkness, Bordner, hoon Kim,
  Kuhlen, Xu, Goldbaum, Hummels, Kritsuk, Tasker, Skory, Simpson, Hahn, Oishi,
  So, Zhao, Cen, \& Li}]{Bryan14}
Bryan, G.~L., Norman, M.~L., O'Shea, B.~W., Abel, T., Wise, J.~H., Turk, M.~J.,
  Reynolds, D.~R., Collins, D.~C., Wang, P., Skillman, S.~W., Smith, B.,
  Harkness, R.~P., Bordner, J., hoon Kim, J., Kuhlen, M., Xu, H., Goldbaum, N.,
  Hummels, C., Kritsuk, A.~G., Tasker, E., Skory, S., Simpson, C.~M., Hahn, O.,
  Oishi, J.~S., So, G.~C., Zhao, F., Cen, R., \& Li, Y. 2014, ApJS, 211, 19

\bibitem[{Carroll-Nellenback {et~al.}(2012)Carroll-Nellenback, Shroyer, Frank,
  \& Ding}]{Carroll12}
Carroll-Nellenback, J., Shroyer, B., Frank, A., \& Ding, C. 2012, ASPC, 459,
  291

\bibitem[{Cox \& Raymond(1985)}]{Cox85}
Cox, D.~P. \& Raymond, J.~C. 1985, ApJ, 298, 651

\bibitem[{Cunningham {et~al.}(2009)Cunningham, Frank, Varni{\`e}re, Mitran, \&
  Jones}]{Cunningham09}
Cunningham, A.~J., Frank, A., Varni{\`e}re, P., Mitran, S., \& Jones, T.~W.
  2009, ApJS, 182, 519

\bibitem[{Dalgarno \& McCray(1972)}]{Dalgarno72}
Dalgarno, A. \& McCray, R.~A. 1972, ARA\&A, 10, 375

\bibitem[{Ferland {et~al.}(1998)Ferland, Korista, Verner, Ferguson, Kingdom, \&
  Verner}]{Ferland98}
Ferland, G.~J., Korista, K.~T., Verner, D.~A., Ferguson, J.~W., Kingdom, J.~B.,
  \& Verner, E.~M. 1998, PASP, 110, 761

\bibitem[{Frank {et~al.}(2014)Frank, Ray, Cabrit, Hartigan, Arce, Bacciotti,
  Bally, Benisty, Eisl{\"o}ffel, Gudel, Lebedev, Nisini, \& Raga}]{Frank14}
Frank, A., Ray, T.~P., Cabrit, S., Hartigan, P., Arce, H.~G., Bacciotti, F.,
  Bally, J., Benisty, M., Eisl{\"o}ffel, J., Gudel, M., Lebedev, S., Nisini,
  B., \& Raga, A. 2014, in Protostars and Planets VI (University of Arizona
  Press), 415

\bibitem[{Fryxell {et~al.}(2000)Fryxell, Olson, Ricker, Timmes, Zingale, Lamb,
  MacNeice, Rosneer, Truran, \& Tufo}]{Fryxell00}
Fryxell, B., Olson, K., Ricker, P., Timmes, F.~X., Zingale, M., Lamb, D.~Q.,
  MacNeice, P., Rosneer, R., Truran, J.~W., \& Tufo, H. 2000, ApJS, 131, 273

\bibitem[{Gaspari {et~al.}(2011)Gaspari, Melioli, Brighenti, \&
  D'Ercole}]{Gaspari11}
Gaspari, M., Melioli, C., Brighenti, F., \& D'Ercole, A. 2011, MNRAS, 411, 349

\bibitem[{Hansen {et~al.}(2015{\natexlab{a}})Hansen, Frank, \&
  Hartigan}]{Hansen15a}
Hansen, E.~C., Frank, A., \& Hartigan, P. 2015{\natexlab{a}}, ApJ, 800, 41H

\bibitem[{Hansen {et~al.}(2017)Hansen, Frank, Hartigan, \& Lebedev}]{Hansen17}
Hansen, E.~C., Frank, A., Hartigan, P., \& Lebedev, S.~V. 2017, ApJ, 837, 143

\bibitem[{Hansen {et~al.}(2015{\natexlab{b}})Hansen, Frank, Hartigan, \&
  Yirak}]{Hansen15b}
Hansen, E.~C., Frank, A., Hartigan, P., \& Yirak, K. 2015{\natexlab{b}}, HEDP,
  17, 135

\bibitem[{Hartigan {et~al.}(2011)Hartigan, Frank, Foster, Wilde, Douglas,
  Rosen, Coker, Blue, \& Hansen}]{Hartigan11}
Hartigan, P., Frank, A., Foster, J.~M., Wilde, B.~H., Douglas, M., Rosen,
  P.~A., Coker, R.~F., Blue, B.~E., \& Hansen, J.~F. 2011, ApJ, 736, 29

\bibitem[{Hartigan \& Wright(2015)}]{Hartigan15}
Hartigan, P. \& Wright, A. 2015, ApJ, 811, 12

\bibitem[{Heathcote {et~al.}(1996)Heathcote, Morse, Hartigan, Reipurth,
  Schwartz, Bally, \& Stone}]{Heathcote96}
Heathcote, S., Morse, J.~A., Hartigan, P., Reipurth, B., Schwartz, R.~D.,
  Bally, J., \& Stone, J.~M. 1996, AJ, 112, 1141

\bibitem[{Keenan {et~al.}(1996)Keenan, Aller, Bell, Hyung, McKenna, \&
  Ramsbottom}]{Keenan96}
Keenan, F.~P., Aller, L.~H., Bell, K.~L., Hyung, S., McKenna, F.~C., \&
  Ramsbottom, C.~A. 1996, MNRAS, 281, 1073

\bibitem[{Li {et~al.}(2015)Li, Bryan, Ruszkowski, Voit, O'Shea, \&
  Donahue}]{Li15}
Li, Y., Bryan, G.~L., Ruszkowski, M., Voit, M.~G., O'Shea, B.~W., \& Donahue,
  M. 2015, ApJ, 811, 73

\bibitem[{Mazzotta {et~al.}(1998)Mazzotta, Mazzitelli, Colafrancesco, \&
  Vittorio}]{Mazzotta98}
Mazzotta, P., Mazzitelli, G., Colafrancesco, S., \& Vittorio, N. 1998, A\&AS,
  133, 403

\bibitem[{Mendoza \& Zeippen(1982)}]{Mendoza82}
Mendoza, C. \& Zeippen, C.~J. 1982, MNRAS, 198, 127

\bibitem[{Raga {et~al.}(1997)Raga, Mellema, \& Lundqvist}]{Raga97}
Raga, A.~C., Mellema, G., \& Lundqvist, P. 1997, ApJS, 109, 517

\bibitem[{Raga {et~al.}(2000)Raga, Navarro-Gonz{\'a}lez, \&
  Villagr{\'a}n-Muniz}]{Raga00}
Raga, A.~C., Navarro-Gonz{\'a}lez, R., \& Villagr{\'a}n-Muniz, M. 2000, RMxAA,
  36, 67

\bibitem[{Raymond(1979)}]{Raymond79}
Raymond, J.~C. 1979, ApJS, 39, 1

\bibitem[{Rosdahl {et~al.}(2013)Rosdahl, Blaizot, Aubert, Stranex, \&
  Teyssier}]{Rosdahl13}
Rosdahl, J., Blaizot, J., Aubert, D., Stranex, T., \& Teyssier, R. 2013, MNRAS,
  436, 2188

\bibitem[{Sutherland \& Dopita(1993)}]{Sutherland93}
Sutherland, R.~S. \& Dopita, M.~A. 1993, ApJS, 88, 253

\bibitem[{Te{\c s}ileanu {et~al.}(2014)Te{\c s}ileanu, Matsakos, Massaglia,
  Trussoni, Mignone, Vlahakis, Tsinganos, Stute, Cayatte, Sauty, Stehl{\'e}, \&
  Chi{\'e}ze}]{Tesileanu14}
Te{\c s}ileanu, O., Matsakos, T., Massaglia, S., Trussoni, E., Mignone, A.,
  Vlahakis, N., Tsinganos, K., Stute, M., Cayatte, V., Sauty, C., Stehl{\'e},
  C., \& Chi{\'e}ze, J.-P. 2014, A\&A, 562, 117

\bibitem[{Te{\c s}ileanu {et~al.}(2008)Te{\c s}ileanu, Mignone, \&
  Massaglia}]{Tesileanu08}
Te{\c s}ileanu, O., Mignone, A., \& Massaglia, S. 2008, A\&A, 488, 429

\bibitem[{Teyssier(2002)}]{Teyssier02}
Teyssier, R. 2002, A\&A, 385, 337

\bibitem[{Verner \& Ferland(1996)}]{Verner96}
Verner, D.~A. \& Ferland, G.~J. 1996, ApJS, 103, 467V

\bibitem[{Yirak {et~al.}(2010)Yirak, Frank, \& Cunningham}]{Yirak10}
Yirak, K., Frank, A., \& Cunningham, A.~J. 2010, ApJ, 722, 412

\bibitem[{Ziegler(2016)}]{Ziegler16}
Ziegler, U. 2016, A\&A, 586, 82

\end{thebibliography}
\end{document}